\documentclass[letterpaper]{article} 
\usepackage{aaai25}  
\usepackage{times}  
\usepackage{helvet}  
\usepackage{courier}  
\usepackage[hyphens]{url}  
\usepackage{graphicx} 
\usepackage{natbib}  
\usepackage{caption} 
\usepackage{algorithm}
\usepackage{algorithmic}
\usepackage{amsmath}  
\usepackage{amssymb}  
\usepackage{booktabs}
\usepackage{newfloat}
\usepackage{listings}
\DeclareCaptionStyle{ruled}{labelfont=normalfont,labelsep=colon,strut=off} 
\floatstyle{ruled}
\newfloat{listing}{tb}{lst}{}
\floatname{listing}{Listing}
\title{Plug-and-Play Tri-Branch Invertible Block for Image Rescaling}

\author {
    Jingwei Bao\textsuperscript{\rm 1,\rm 2},
    Jinhua Hao\textsuperscript{\rm 2}\thanks{Corresponding author.}\thanks{Project leader.},
    Pengcheng Xu\textsuperscript{\rm 2}\footnotemark[1]\footnotemark[2],
    Ming Sun\textsuperscript{\rm 2},
    Chao Zhou\textsuperscript{\rm 2},
    Shuyuan Zhu\textsuperscript{\rm 1}\footnotemark[1]
}
\affiliations {
    \textsuperscript{\rm 1} University of Electronic Science and Technology of China, Chengdu, China\\
    \textsuperscript{\rm 2} Kuaishou Technology, Beijing, China\\
    jwbao@std.uestc.edu.cn, \{haojinhua, xupengcheng, sunming03, zhouchao\}@kuaishou.com, eezsy@uestc.edu.cn
}

\begin{document}

\maketitle

\begin{abstract}
High-resolution (HR) images are commonly downscaled to low-resolution (LR) to reduce bandwidth, followed by upscaling to restore their original details. Recent advancements in image rescaling algorithms have employed invertible neural networks (INNs) to create a unified framework for downscaling and upscaling, ensuring a one-to-one mapping between LR and HR images. Traditional methods, utilizing dual-branch based vanilla invertible blocks, process high-frequency and low-frequency information separately, often relying on specific distributions to model high-frequency components. However, processing the low-frequency component directly in the RGB domain introduces channel redundancy, limiting the efficiency of image reconstruction. To address these challenges, we propose a plug-and-play tri-branch invertible block (T-InvBlocks) that decomposes the low-frequency branch into luminance (Y) and chrominance (CbCr) components, reducing redundancy and enhancing feature processing. Additionally, we adopt an all-zero mapping strategy for high-frequency components during upscaling, focusing essential rescaling information within the LR image. Our T-InvBlocks can be seamlessly integrated into existing rescaling models, improving performance in both general rescaling tasks and scenarios involving lossy compression. Extensive experiments confirm that our method advances the state of the art in HR image reconstruction.
\begin{links}
\link{Code}{https://github.com/Jingwei-Bao/T-InvBlocks}
\end{links}
\end{abstract}

\section{Introduction}
\begin{figure}[t]
    \centering
    \includegraphics[width=\linewidth]{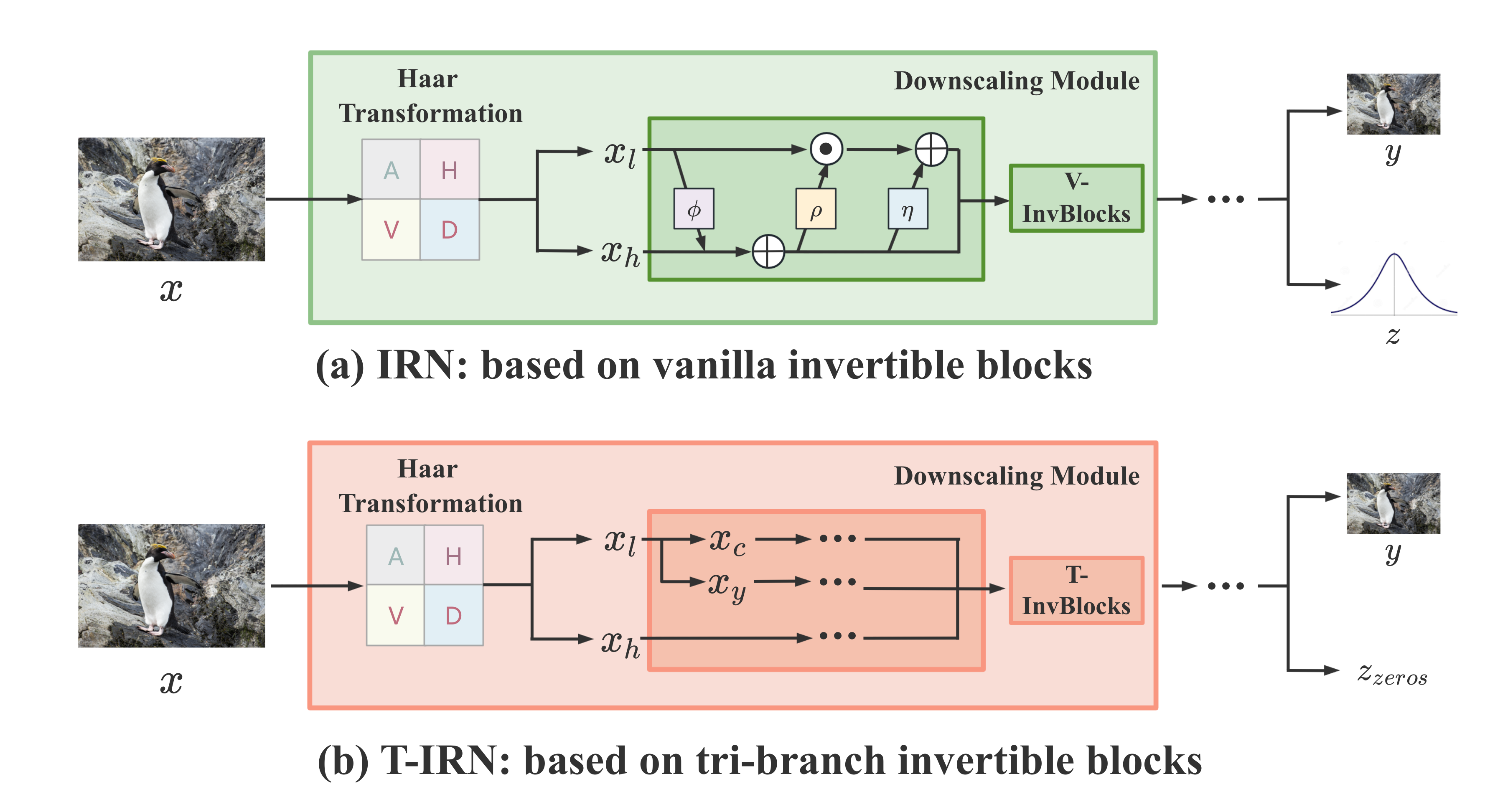}
    \caption{Comparison of IRN and T-IRN: (a) Vanilla invertible block in IRN processing low-frequency components in RGB. (b) Tri-branch invertible block in T-IRN processing low-frequency components in YCbCr, with separate branches for Y and CbCr.}
    \label{fig:dual}
\end{figure}

High-resolution (HR) images are extensively distributed across online networks. To optimize display and conserve bandwidth and storage, HR images are typically downscaled to low-resolution (LR) versions that preserve visual integrity. Consequently, upscaling is essential to restore LR images to their original size and detail. Super-resolution methods~\cite{dong2015image,dai2019second,chen2024cassr,qu2025xpsr,qin2025new} enhance image resolution but typically assume a predetermined, non-learnable downscaling operator. Recent studies~\cite{kim2018task,xiao2020invertible,guo2022invertible} have aimed to jointly optimize both downscaling and upscaling processes. IRN~\cite{xiao2020invertible} models the transformation between HR and LR images as a bijective mapping using an invertible neural network (INN) based on a vanilla invertible block (V-InvBlock) based on dual-branch. This method employs the Haar wavelet transform~\cite{lienhart2002extended} to decompose the image into low- and high-frequency components, which are processed by an INN. The network enforces high-frequency components to follow a case-agnostic normal distribution, preserving as much information as possible, as shown in Figure~\ref{fig:dual}(a). Additionally, LR images are often compressed to reduce bandwidth and storage~\cite{son2021enhanced}, but suffer quality degradation~\cite{mo2025oapt,zhu2024cpga}. To address this, SAIN~\cite{yang2023self} builds on IRN by introducing an asymmetric model structure and incorporating simulated compression to enhance robustness. SAIN employs a learnable Gaussian mixture model (GMM) to sample and fit the lost high-frequency information during upscaling, based on the observed high-frequency distribution during downscaling.

Despite the widespread use of dual-branch invertible neural networks for image rescaling, we believe there is room for optimization. The V-InvBlock superficially addresses the interaction between high- and low-frequency components, without fully exploring the integration of information across different domains to establish a more effective bijective mapping between LR and HR. The low-frequency components, preserved as the LR output, are the only intermediate result retained throughout the process and are crucial for restoring image details during upscaling, warranting further processing. Moreover, IRN’s ablation studies~\cite{xiao2023invertible} reveal that using an all-zero tensor instead of sampling from a Gaussian distribution can yield comparable or even superior upscaling results.

To address these issues, we propose a tri-branch invertible block that further decomposes the low-frequency branch into separate luminance (Y) and chrominance (CbCr) components. This approach is inspired by the understanding that luminance and chrominance belong to different domains of image information, each with distinct characteristics. By converting the low-frequency information into YCbCr color space and processing these components separately, the model reduces inter-channel redundancy, enhancing its ability to interact with and process different domains of image information more effectively. This approach improves the network's capacity to accurately capture and reconstruct critical image details, thereby strengthening the overall quality of the rescaling transformation between LR and HR. Additionally, during upscaling in training process, we replace the high-frequency sampling operation with an all-zero tensor, a strategy aimed at concentrating the reconstruction task on the inherent details of the LR image itself, rather than relying on randomly sampled high-frequency information, as shown in Figure~\ref{fig:dual}(b). This approach leverages the essential content within the LR image, which our experiments suggest leads to more accurate and robust HR reconstruction. We integrated our proposed blocks with the strategy into both IRN and SAIN, resulting in T-IRN and T-SAIN, which effectively demonstrate the plug-and-play nature of the T-InvBlock. This straightforward replacement of existing components without altering the overall network architecture allowed us to enhance performance across both general image rescaling tasks and scenarios involving lossy compression, highlighting the effectiveness and robustness of our approach.

Our main contributions are as follows:
\begin{itemize}
    \item We introduce T-InvBlock, which separates the low-frequency RGB information into luminance (Y) and chrominance (CbCr) components. This reduces inter-channel redundancy and allows specialized processing of Y and CbCr, enhancing the model's effectiveness in mapping LR to HR.

    \item We introduce a training strategy that replaces high-frequency sampling with an all-zero tensor, concentrating the information within the LR image, thereby reducing the risk of inconsistencies introduced by distribution sampling and enhancing the accuracy of restoration.

    \item We seamlessly integrate our block into existing IRN and SAIN models for both general rescaling and scenarios involving lossy compression of LR images. Experiments confirm the plug-and-play T-InvBlock's effectiveness, and robustness, with improvements in performance.

\end{itemize}

\begin{figure*}[ht]
    \centering
    \includegraphics[width=\linewidth]{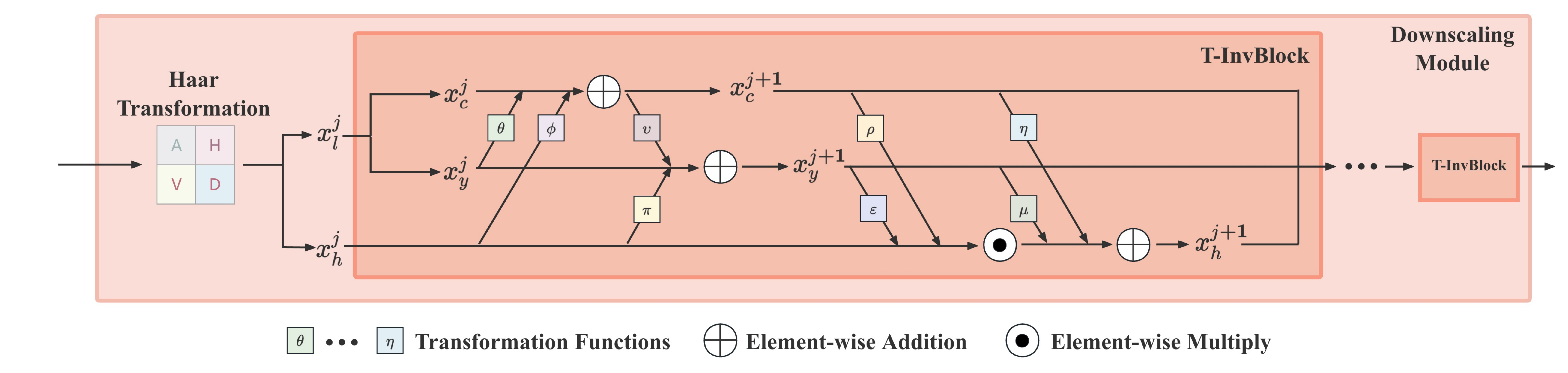}
    \caption{Architecture of tri-branch invertible block(T-InvBlock)}
    \label{fig:T-block}
\end{figure*}
\section{Related Work}
\paragraph{Image Rescaling.}
Super-resolution (SR)~\cite{dong2015image,lim2017enhanced,zhang2018residual} aims to reconstruct HR images from LR inputs, which are typically downscaled using algorithm like Bicubic interpolation, leading to the loss of high-frequency details. In contrast, image rescaling~\cite{kim2018task,li2018learning,sun2020learned} jointly optimizes both downscaling and upscaling processes to better preserve image quality. Notably, IRN~\cite{xiao2020invertible} employs invertible neural networks (INNs) to model rescaling as a bijective transformation, where the downscaling and upscaling are not independent operations but are instead mutually inverse processes within the same model. The high-frequency components of the HR image are embedded into a case-agnostic latent space during downscaling, ensuring that as much information as possible is retained and can be accurately recovered during upscaling. HCFlow~\cite{liang2021hierarchical} extends this by integrating SR and rescaling tasks into a single framework, using conditional distributions for high-frequency details. BAIRNet~\cite{pan2022towards} and IARN~\cite{pan2023effective} introduce novel techniques like subpixel splitting and preemptive channel splitting to handle unconventional scaling factors. AIDN~\cite{xing2023scale} further generalizes this approach by allowing for arbitrary scaling factors. In practical applications, downscaling is often paired with image compression, which can significantly degrade the quality of HR image reconstruction due to the loss of critical details in the compression process. To address this, SAIN~\cite{yang2023self} proposes an asymmetric invertible framework that simulates compression during downscaling and employs actual image codecs (e.g., JPEG) during upscaling, enhancing the model's robustness to lossy compression when reconstructing HR images.
\paragraph{Invertible Neural Networks.}
Invertible Neural Networks (INNs), widely used in image rescaling, are derived from flow-based generative models. These networks ensure bijective mappings between source and target domains, allowing for precise transformations. Through normalizing flows~\cite{rezende2015variational,kobyzev2020normalizing,liu2021invertible,quan2024enhancing}, INNs convert high-dimensional data, like images, into simple latent distributions such as Gaussian, enabling efficient computation of the Jacobian determinant and optimization of posterior probabilities via maximum likelihood estimation (MLE).

In the context of image rescaling tasks, the IRN model~\cite{xiao2020invertible} leverages a dual-branch invertible block (V-InvBlock) to process the HR image $x \in \scalebox{0.85}{$\mathbb{R}^{N \times H \times W \times C}$}$, decomposing it into low-frequency components $x_l \in \scalebox{0.85}{$\mathbb{R}^{N \times H \times W \times C/4}$}$ and high-frequency components $x_h \in \scalebox{0.85}{$\mathbb{R}^{N \times H \times W \times 3C/4}$}$ using the Haar wavelet transform. The low-frequency branch of IRN is based on the additive coupling layer introduced by~\cite{dinh2014nice}, while the high-frequency branch utilizes a more expressive affine coupling layer as proposed by~\cite{dinh2016density}. The transformations applied in IRN are formalized as follows:
\begin{equation}
\begin{aligned}
    &x_{l}^{j+1} = x_{l}^{j} + \phi(x_{h}^{j})~, \\
    &x_{h}^{j+1} = x_{h}^{j} \odot \exp (\rho(x_{l}^{j+1})) + \eta(x_{l}^{j+1})~, \\
    &x_{h}^{j} = (x_{h}^{j+1} - \eta(x_{l}^{j+1})) \odot \exp (-\rho(x_{l}^{j+1}))~, \\
    &x_{l}^{j} = x_{l}^{j+1} - \phi(x_{h}^{j})~.
\end{aligned}
\label{eq:IRNinvblock}
\end{equation}

In the $j$-th block, $x_l^j$ represents the low-frequency component, while $x_h^j$ denotes the high-frequency component. These transformations ensure that the model effectively captures the essential details in both the low- and high-frequency domains, while maintaining the invertibility of the overall network. The transformation functions $\phi(\cdot)$, $\eta(\cdot)$, and $\rho(\cdot)$ are parameterized by densely connected convolutional blocks~\citep{wang2018esrgan}.

Additionally, SAIN~\cite{yang2023self} builds upon this V-InvBlock by introducing the Enhanced Invertible Block (E-InvBlock), which adds some transformation functions specifically for the downscaling module. However, the compression simulation module in SAIN retains the design of IRN, maintaining the V-InvBlocks with dual-branch structure, as shown in Figure~\ref{fig:T-INN}(a).

\section{Methodology}
\subsection{Preliminaries}
In image rescaling tasks, dual-branch invertible neural networks, which process high and low-frequency components separately after Haar wavelet transformation, are widely adopted. IRN~\cite{xiao2020invertible} splits HR images $x$ into low- and high-frequency components $[x_l, x_h]$, learning an invertible mapping $[x_l, x_h] \leftrightarrow [y, z]$, where $y$ is the LR output and $z \sim \mathcal{N}(0,1)$ is a case-agnostic variable. SAIN enhances robustness against image compression by establishing an asymmetric framework. Here, the downscaling process is modeled by $f$, which outputs a high-quality LR image $y$, and the compression is simulated by $g$. The combined operation $f \circ g$ then maps $[x_l, x_h]$ to $[\hat{y}, \hat{z}]$, where $\hat{y}$ is the simulated compressed LR image. When compression artifacts, introduced by $\varepsilon$, degrade $y$ to $y^{\prime}$, the reverse pass first applies the inverse compression function $g^{-1}$ to $y^{\prime}$, recovering it to $y_r$. Then, the reverse pass reconstructs the HR image by mapping $[y_r, z] \rightarrow x^\prime$, with $z$ modeled by a learnable GMM, as shown in Figure~\ref{fig:T-INN}.

In this work, we propose our plug-and-play T-InvBlock for enhanced image rescaling. After applying the Haar wavelet transform, the low-frequency component $x_l$ is converted to YCbCr color space and decomposed into $x_y$ (luminance) and $x_c$ (chrominance) channels, which are then processed through the T-InvBlock’s three branches. This structure reduces inter-channel redundancy and allows for more targeted processing of luminance and chrominance information.  At the final stage of downscaling, $x_y$ and $x_c$ are concatenated, converted back to the original color space, and output as the LR image $y$. The model learns an invertible mapping $[x_y, x_c, x_h] \leftrightarrow [y, z]$, where during upscaling, $z$ is set to an all-zero tensor. This approach shifts the responsibility of encoding high-frequency details from a statistical distribution (e.g., fixed Gaussian in IRN or learnable GMM in SAIN) to the LR image itself, concentrating relevant information within the LR representation. By seamlessly integrating the T-InvBlock into the IRN and SAIN frameworks, we developed T-IRN and T-SAIN, which, as illustrated in Figure~\ref{fig:dual}(b) and Figure~\ref{fig:T-INN}(b), demonstrated superior results across various rescaling tasks, including those involving lossy compression, thereby proving the effectiveness and versatility of our approach.

\subsection{Tri-branch Invertible Block}
Previous image rescaling methods using dual-branch invertible neural networks process low-frequency ($x_l$) and high-frequency ($x_h$) components from Haar wavelet transformation separately. We propose an improvement by converting $x_l$ to YCbCr, decomposing it into luminance (Y) and chrominance (CbCr) components, and processing these through independent branches for more effective interaction and analysis. Given that $x_l$ is the only element retained throughout the rescaling process, it warrants deeper processing.

Converting low-frequency information into the YCbCr color space and segregating the luminance (Y) from the chrominance (CbCr) components, rather than directly processing low-frequency RGB data, offers several significant advantages. This conversion effectively reduces inter-channel redundancy~\cite{jain1989fundamentals,poynton2012digital}, thereby enhancing the overall efficiency of information representation and allowing the model to concentrate more precisely on the critical features that influence image quality. Furthermore, the distinct operational domains of luminance and chrominance facilitate the application of more tailored processing techniques for each component, which can lead to optimized performance outcomes. This strategy is especially advantageous in scenarios involving JPEG-compressed LR images, as JPEG compression inherently operates in the YCbCr domain~\cite{wallace1991jpeg}. By aligning the model's processing framework with this compression standard, the model is better equipped to adapt to compression artifacts, ultimately leading to superior reconstruction quality. As a result, we developed a novel downscaling module based on the T-InvBlock architecture, as illustrated in Figure~\ref{fig:T-block}.

Based on Equation~\ref{eq:IRNinvblock}, the T-InvBlock can be formulated as follows:
\begin{equation}
    \begin{aligned}
    &x_{c}^{j+1} = x_{c}^{j} + \phi(x_{h}^{j}) + \theta(x_{y}^{j}), \\
    &x_{y}^{j+1} = x_{y}^{j} + \nu(x_{c}^{j+1}) + \pi(x_{h}^{j}), \\
    &x_{h}^{j+1} = x_{h}^{j} \odot \exp(\rho(x_{c}^{j+1}) + \varepsilon(x_{y}^{j+1}))\\
    &\qquad \quad + \eta(x_{c}^{j+1}) + \mu(x_{y}^{j+1}).
    \label{eq:TIRNinvblock-down}
    \end{aligned}
\end{equation}
The upscaling process is:
\begin{equation}
    \begin{aligned}
    &x_{h}^{j} = (x_{h}^{j+1} - \eta(x_{c}^{j+1}) - \mu(x_{y}^{j+1})) \\
    &\qquad \quad \odot \exp(-\rho(x_{c}^{j+1}) - \varepsilon(x_{y}^{j+1})), \\
    &x_{y}^{j} = x_{y}^{j+1} - \nu(x_{c}^{j+1}) - \pi(x_{h}^{j}), \\
    &x_{c}^{j} = x_{c}^{j+1} - \phi(x_{h}^{j}) - \theta(x_{y}^{j}).
    \end{aligned}
\label{eq:TIRNinvblock-up}
\end{equation}

In the $j$-th T-InvBlock, $x_{c}^j$ represents the chrominance components, and $x_y^j$ denotes the luminance component of the low-frequency information, while $x_h^j$ represents the high-frequency component. The transformation functions $\phi(\cdot)$, $\theta(\cdot)$, $\nu(\cdot)$, $\pi(\cdot)$, $\rho(\cdot)$, $\varepsilon(\cdot)$, $\eta(\cdot)$, and $\mu(\cdot)$ are parameterized by densely connected convolutional blocks~\citep{wang2018esrgan}, similar to those used in the V-InvBlock, and are designed to handle the interactions and updates across the different components within the block.
\subsection{All-zero mapping Strategy}
We propose an all-zero mapping strategy for modeling high-frequency components during upscaling, aiming to enhance the performance of image rescaling tasks by more effectively utilizing the information within the LR image. This approach is inspired by two key observations. First, experiments with IRN show that using an all-zero tensor for the initial high-frequency component during upscaling can achieve similar or even better HR reconstruction performance than using tensors sampled from a fixed Gaussian distribution. This suggests that the precise modeling of high-frequency components may not be as crucial to final image quality as previously thought. Second, although the SAIN model employs a GMM to learn the latent distribution of high-frequency components, the official weights show that the GMM actually learns a near-zero mean unimodal distribution. This further indicates that high-frequency information in practice does not exhibit complex distribution characteristics, implying that a simple all-zero tensor may suffice for effective HR image reconstruction.

Building on these insights, we adopt an all-zero tensor, i.e., $z=0$, for high-frequency components during upscaling in models' training and testing processes. This approach shifts the responsibility for representing high-frequency details to the LR image itself, which we hypothesize could offer a more deterministic and content-specific basis for reconstruction compared to abstract statistical models like fixed Gaussian distributions or learnable GMM, which introduce randomness that may not always align with the actual content of the image. Our experiments suggest that by mapping high-frequency information to an all-zero tensor, the model can more effectively utilize the inherent information within the LR image, leading to more accurate and robust HR reconstruction. This strategy leverages the essential content already present in the LR image, resulting in enhanced performance in the upscaling process.
\begin{table*} [t]
	\centering
	\small
    \setlength{\tabcolsep}{1mm}
	\begin{tabular}{c|c|c|c|c|c|c|c}
		\hline
		Downscaling \& Upscaling  &  Scale & Param  &  Set5  &  Set14  &  BSD100  &  Urban100  &  DIV2K \\
		
		\hline  
		\hline
		Bicubic \& Bicubic & 2$\times$ & / & 33.66 / 0.9299 & 30.24 / 0.8688 & 29.56 / 0.8431 & 26.88 / 0.8403 & 31.01 / 0.9393 \\
		
		\hline
		Bicubic \& SRCNN \citep{dong2015image} & 2$\times$ & 57.3K &  36.66 / 0.9542  &  32.45 / 0.9067  &  31.36 / 0.8879  &  29.50 / 0.8946 &   35.60 / 0.9663  \\
		
		\hline
		Bicubic \& EDSR \citep{lim2017enhanced} & 2$\times$ &  40.7M & 38.20 / 0.9606  &  34.02 / 0.9204  &  32.37 / 0.9018  &  33.10 / 0.9363 &  35.12 / 0.9699 \\
		
		\hline
		Bicubic \& RDN \citep{zhang2018residual} & 2$\times$ & 22.1M &  38.24 / 0.9614  &  34.01 / 0.9212  &  32.34 / 0.9017  &  32.89 / 0.9353 &   --  \\
		
		\hline
		Bicubic \& RCAN \citep{zhang2018image} & 2$\times$ & 15.4M &  38.27 / 0.9614  &  34.12 / 0.9216  &  32.41 / 0.9027  &  33.34 / 0.9384  &  --  \\
		
		\hline
		Bicubic \& SAN \citep{dai2019second} & 2$\times$ & 15.7M &  38.31 / 0.9620  &  34.07 / 0.9213  &  32.42 / 0.9028  &  33.10 / 0.9370  &  --  \\
		
		\hline
		TAD \& TAU \citep{kim2018task} & 2$\times$ & -- & 38.46 /  --   & 35.52 /  --  & 36.68 /  --   & 35.03 /  --  & 39.01 /  --  \\
		
		\hline
		CNN-CR \& CNN-SR \citep{li2018learning} & 2$\times$ & -- & 38.88 / -- & 35.40 / -- & 33.92 / -- & 33.68 / -- & --\\
		
		\hline
		CAR \& EDSR \citep{sun2020learned} & 2$\times$ & 51.1M & 38.94 / 0.9658 & 35.61 / 0.9404 & 33.83 / 0.9262 & 35.24 / 0.9572 & 38.26 / 0.9599 \\
		
		\hline
		IRN\citep{xiao2020invertible} & 2$\times$ & 1.67M & {43.99} / {0.9871} & {40.79} / {0.9778} & {41.32} / {0.9876} & {39.92} / {0.9865} & {44.32} / {0.9908} \\
		\hline
		T-IRN (Ours) & 2$\times$ & 1.57M & \textbf{44.86} / \textbf{0.9883} & \textbf{41.70} / \textbf{0.9809} & \textbf{42.68} / \textbf{0.9913} & \textbf{41.05} / \textbf{0.9899} & \textbf{45.46} / \textbf{0.9932}

 \\
		\hline  
		\hline
		Bicubic \& Bicubic & 4$\times$ & / & 28.42 / 0.8104 & 26.00 / 0.7027 & 25.96 / 0.6675 & 23.14 / 0.6577 & 26.66 / 0.8521 \\
		
		\hline
		Bicubic \& SRCNN \citep{dong2015image} & 4$\times$ & 57.3K &  30.48 / 0.8628  &  27.50 / 0.7513  &  26.90 / 0.7101  &  24.52 / 0.7221 &   --  \\
		
		\hline
		Bicubic \& EDSR \citep{lim2017enhanced} & 4$\times$ & 43.1M & 32.62 / 0.8984 & 28.94 / 0.7901 & 27.79 / 0.7437 & 26.86 / 0.8080 & 29.38 / 0.9032 \\

		\hline
		Bicubic \& RDN \citep{zhang2018residual} & 4$\times$ & 22.3M &  32.47 / 0.8990  &  28.81 / 0.7871  &  27.72 / 0.7419  &  26.61 / 0.8028 &   --  \\
		
		\hline
		Bicubic \& RCAN \citep{zhang2018image} & 4$\times$ & 15.6M & 32.63 / 0.9002 & 28.87 / 0.7889 & 27.77 / 0.7436 & 26.82 / 0.8087 & 30.77 / 0.8460 \\
		
		\hline
		Bicubic \& ESRGAN \citep{wang2018esrgan} & 4$\times$ & 16.3M & 32.74 / 0.9012 & 29.00 / 0.7915 & 27.84 / 0.7455 & 27.03 / 0.8152 & 30.92 / 0.8486 \\
		
		\hline
		Bicubic \& SAN \citep{dai2019second} & 4$\times$ & 15.7M &  32.64 / 0.9003  &  28.92 / 0.7888  &  27.78 / 0.7436  &  26.79 / 0.8068  &  --  \\
		
		\hline
		TAD \& TAU \citep{kim2018task} & 4$\times$ & -- & 31.81 /  --  & 28.63 /  --   & 28.51 /  --   & 26.63 /  --   & 31.16 /  --   \\
		
		\hline
		CAR \& EDSR \citep{sun2020learned} & 4$\times$ & 52.8M & 33.88 / 0.9174 & 30.31 / 0.8382 & 29.15 / 0.8001 & 29.28 / 0.8711 & 32.82 / 0.8837 \\
		
		\hline
		IRN\citep{xiao2020invertible} & 4$\times$ & 4.35M & 36.19 / 0.9451 & {32.67} / \textbf{0.9015} & {31.64} / {0.8826} & \textbf{31.41} / \textbf{0.9157} & {35.07} / {0.9318} \\
        \hline
        T-IRN (Ours) & 4$\times$ & 4.67M & \textbf{36.29} / \textbf{0.9452} & \textbf{32.70} / {0.9003} & {31.64} / \textbf{0.8837} & {31.19} / {0.9132} & \textbf{35.10} / \textbf{0.9328}\\
        \hline    
	\end{tabular}
    \caption{Quantitative evaluation results (PSNR / SSIM) of different downscaling and upscaling methods for image reconstruction on benchmark datasets: Set5, Set14, BSD100, Urban100, and DIV2K validation set.}
    \label{tab:quantitative results T-IRN}
\end{table*}
\begin{figure}[t]
    \centering
    \includegraphics[width=\linewidth]{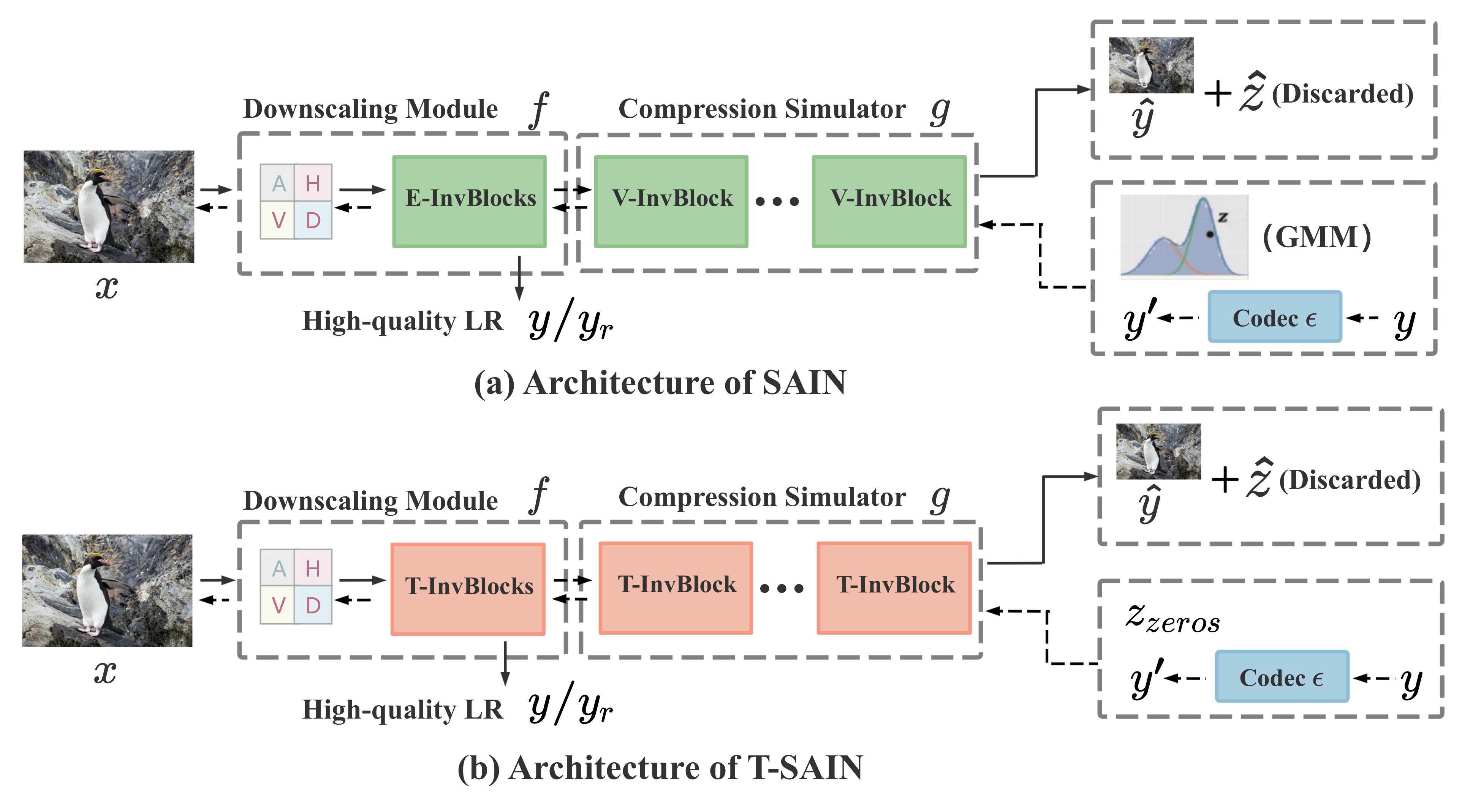}
    \caption{Architecture of SAIN and T-SAIN}
    \label{fig:T-INN}
\end{figure}
\subsection{T-IRN and T-SAIN}
We enhanced the IRN and SAIN models by seamlessly integrating our newly designed T-InvBlock. This integration highlights the plug-and-play nature of the T-InvBlock, allowing us to replace the V-InvBlock or E-InvBlock in the original models without altering the overall network architecture. By introducing the T-InvBlock and implementing our all-zero mapping strategy, we developed the T-IRN and T-SAIN models. These enhanced models not only preserve the robustness and effectiveness of the original frameworks but also deliver significant performance improvements across various image rescaling tasks.

For T-IRN, the invertible blocks in the original downscaling module are substituted with T-InvBlocks. The all-zero mapping strategy eliminates the need for mapping high-frequency components to a distribution, making the original distribution matching loss $\mathcal{L}_\text{distr}$ unnecessary. Thus, the T-IRN training objective, inherited from IRN, omits this term and is expressed as follows:
\begin{align}
    \mathcal{L} = \lambda_1 \mathcal{L}_{hr} + \lambda_2 \mathcal{L}_{lr}.
    \label{eq:TIRN_objective_function}
\end{align}
Here, $\mathcal{L}_{hr}$ ensures that the HR image $x$ can be accurately restored from the model-downscaled LR image $y$ using an all-zero tensor $z$, while $\mathcal{L}_{lr}$ maintains the perceptual quality of the LR images by matching them to those produced by Bicubic interpolation. The terms $\lambda_1$ and $\lambda_2$ are the respective loss weights.

For T-SAIN, we retained the asymmetric model framework and reconstructed the downscaling module $f$ and compression simulator $g$ using T-InvBlocks. Since SAIN does not use distribution-related guidance loss for the GMM during training, we adopted the same loss function as SAIN and replaced the GMM with the all-zero mapping strategy, as shown in Figure~\ref{fig:T-INN}(b).
\begin{table*} [t]
	\centering
	\small  
    \setlength{\tabcolsep}{1.0mm} 
	
	\begin{tabular}{c|c|c|c|c|c|c|c}
		\hline
		Downscaling \& Upscaling & Scale & Param & JPEG QF=30 & JPEG QF=50 & JPEG QF=70 & JPEG QF=80 & JPEG QF=90 \\
		\hline  
		\hline
		Bicubic \& Bicubic & $\times 2$ & / & 29.38 / 0.8081 & 30.19 / 0.8339 & 30.91 / 0.8560 & 31.38 / 0.8703 & 31.96 / 0.8878 \\
		\hline
		Bicubic \& SRCNN \citep{dong2015image} & $\times 2$ & 57.3K & 28.01 / 0.7872 & 28.69 / 0.8154 & 29.43 / 0.8419 & 30.01 / 0.8610 & 30.88 / 0.8878  \\
		\hline
		Bicubic \& EDSR \citep{lim2017enhanced} & $\times 2$ & 40.7M & 28.92 / 0.7947 & 29.93 / 0.8257 & 31.01 / 0.8546 & 31.91 / 0.8753 & 33.44 / 0.9052 \\
		\hline
		Bicubic \& RDN \citep{zhang2018residual} & $\times 2$ & 22.3M & 28.95 / 0.7954 & 29.96 / 0.8265 & 31.02 / 0.8549 & 31.91 / 0.8752 & 33.41 / 0.9046\\
		\hline
		Bicubic \& RCAN \citep{zhang2018image} & $\times 2$ & 15.4M & 28.84 / 0.7932 & 29.84 / 0.8245 & 30.94 / 0.8538 & 31.87 / 0.8749 & 33.44 / 0.9052\\
		\hline
		CAR \& EDSR \citep{sun2020learned} & $\times 2$ & 51.1M & 27.83 / 0.7602 & 28.66 / 0.7903 & 29.44 / 0.8165 & 30.07 / 0.8347 & 31.31 / 0.8648\\
		\hline
		IRN \citep{xiao2020invertible} & $\times 2$ & 1.67M & 29.24 / 0.8051 & 30.20 / 0.8342 & 31.14 / 0.8604 & 31.86 / 0.8783 & 32.91 / 0.9023 \\
		\hline
		SAIN \citep{yang2023self} & $\times 2$ & 2.02M & {31.47 / 0.8747} & {33.17 / 0.9082} & {34.73 / 0.9296} & {35.46 / 0.9374} & {35.96 / 0.9419} \\
		\hline 
        T-SAIN (Ours)  & $\times 2$ & 2.09M & \textbf{31.89 / 0.8912} & \textbf{33.71 / 0.9210} & \textbf{35.20 / 0.9384} & \textbf{35.86 / 0.9443} & \textbf{36.30 / 0.9478} \\
        \hline
		\hline
		Bicubic \& Bicubic & $\times 4$ & / & 26.27 / 0.6945 & 26.81 / 0.7140 & 27.28 / 0.7326 & 27.57 / 0.7456 & 27.90 / 0.7618 \\
		\hline
		Bicubic \& SRCNN \citep{dong2015image} & $\times 4$ & 57.3K & 25.49 / 0.6819 & 25.91 / 0.7012 & 26.30 / 0.7206 & 26.55 / 0.7344 & 26.84 / 0.7521 \\
		\hline
		Bicubic \& EDSR \citep{lim2017enhanced} & $\times 4$ & 43.1M & 25.87 / 0.6793 & 26.57 / 0.7052 & 27.31 / 0.7329 & 27.92 / 0.7550 & 28.88 / 0.7889\\
		\hline
		Bicubic \& RDN \citep{zhang2018residual} & $\times 4$ & 22.3M & 25.92 / 0.6819 & 26.61 / 0.7075 & 27.33 / 0.7343 & 27.92 / 0.7556 & 28.84 / 0.7884\\
		\hline
		Bicubic \& RCAN \citep{zhang2018image} & $\times 4$ & 15.6M & 25.77 / 0.6772 & 26.45 / 0.7031 & 27.21 / 0.7311 & 27.83 / 0.7537 & 28.82 / 0.7884\\
		\hline
		Bicubic \& ESRGAN \citep{wang2018esrgan} & $\times 4$ & 16.3M & 25.87 / 0.6803 & 26.58 / 0.7063 & 27.36 / 0.7343 & 27.99 / 0.7568 & 28.98 / 0.7915 \\
		\hline
		CAR \& EDSR \citep{sun2020learned} & $\times 4$ & 52.8M & 25.25 / 0.6610 & 25.76 / 0.6827 & 26.22 / 0.7037 & 26.69 / 0.7214 & 27.91 / 0.7604\\
		\hline
		IRN \citep{xiao2020invertible} & $\times 4$ & 4.35M & 25.98 / 0.6867 & 26.62 / 0.7096 & 27.24 / 0.7328 & 27.72 / 0.7508 & 28.42 / 0.7777 \\
		\hline
		SAIN \citep{yang2023self} & $\times 4$ & 6.49M & {27.90 / 0.7745} & {29.05 / 0.8088} & {29.83 / 0.8272} & {30.13 / 0.8331} & {30.31 / 0.8367} \\
        \hline
        T-SAIN (Ours)  & $\times 4$ & 5.45M & \textbf{28.08 / 0.7893} & \textbf{29.43 / 0.8237} & \textbf{30.34 / 0.8421} & \textbf{30.69 / 0.8483} & \textbf{30.92 / 0.8517} \\

        \hline
	\end{tabular}
    \caption{Quantitative results (PSNR / SSIM) of image rescaling on DIV2K under distortion at different JPEG QFs.}
    \label{tab:quantitative results T-SAIN}
\end{table*}
\section{Experiments}
\subsection{Experimental Setup}
\paragraph{Datasets and Settings.}
We adopt 800 HR images from the widely-used DIV2K training set~\cite{agustsson2017ntire} to train our models. For evaluation, T-IRN is assessed on five standard test sets: Set5~\cite{bevilacqua2012low}, Set14~\cite{zeyde2010single}, BSD100~\cite{martin2001database}, Urban100~\cite{huang2015single}, and the DIV2K validation set~\cite{agustsson2017ntire}. T-SAIN is evaluated on the DIV2K validation set with JPEG compression at QFs (i.e., 30, 50, 70, 80, 90). While quality assessment methods are diverse~\cite{yuan2024ptm,lu2024kvq,xie2024qpt}, following standard image rescaling practices~\cite{xiao2020invertible, yang2023self}, both models are evaluated using Peak Signal-to-Noise Ratio (PSNR) and Structural Similarity Index (SSIM)~\cite{wang2004image}, measured on the Y channel of the YCbCr color space.
\paragraph{Implementation Details.}

For T-IRN image rescaling, we use a single downscaling module (Haar transformation with 3 T-InvBlocks) for $\times2$ scaling, and two modules with an additional T-InvBlock for $\times4$. The $\times2$ model is trained for 800k iterations, halving the learning rate every 100k iterations, while the $\times4$ model is trained for 600k iterations, with the learning rate halved every 40k iterations. Loss weights are set to $\lambda_1 = 1$ and $\lambda_2 = 0.25$ to balance LR and HR loss and enhance HR detail restoration.

For T-SAIN $\times2$ image rescaling, we employ a single downscaling module and a T-InvBlock as a compression simulator, doubling the modules for $\times4$. Both T-SAIN models are trained for 600k iterations, with the learning rate halved every 100k iterations. The training setup follows SAIN~\cite{yang2023self}, including the loss function and JPEG codec $\varepsilon$ with a QF of 75.

In all T-IRN and T-SAIN experiments, the initial learning rate is $2\times10^{-4}$, using $\mathcal{L}_2$ pixel loss for $\mathcal{L}_{lr}$ and $\mathcal{L}_1$ pixel loss for $\mathcal{L}_{hr}$ in RGB space. Input images are cropped to 128$\times$128 and augmented with random flips. We use the Adam optimizer~\cite{kingma2014adam} with $\beta_1=0.9$, $\beta_2=0.999$, and a mini-batch size of 16.

\begin{table}[t]
    \centering
    \small
    \setlength{\tabcolsep}{1mm}
    \begin{tabular}{l|c|c|c|c}
        \hline
        Strategy &Param& Set14 & Urban100 & DIV2K \\
        \hline
        IRN &1.67M& 40.79 & 39.92 & 44.32 \\
        \hline
        IRN (YCbCr) &1.67M& 40.47 & 39.33 & 44.02 \\
        \hline
        IRN (all-zero) &1.67M& 41.37 & 40.55 & 45.18 \\
        \hline
        T-IRN&1.57M& \textbf{41.70} & \textbf{41.05} & \textbf{45.46} \\
        \hline
    \end{tabular}
    \caption{Ablation study on T-IRN: PSNR of different image rescaling strategies for $\times 2$ image reconstruction on Set14, Urban100 and DIV2K}
    \label{tab:ablaction_study_TIRN}
\end{table}
\subsection{Evaluation for T-IRN}
\subsubsection{Quantitative Evaluation.}
We compared T-IRN against two types of methods: (1) approaches based on Bicubic downscaling combined with super-resolution~\citep{dong2015image,lim2017enhanced,zhang2018residual,zhang2018image,wang2018esrgan,dai2019second}, and (2) methods that perform jointly-optimized downscaling and upscaling~\citep{kim2018task,li2018learning,sun2020learned}, especially original IRN~\citep{xiao2020invertible}.

As shown in Table~\ref{tab:quantitative results T-IRN}, T-IRN outperforms state-of-the-art baseline models, including IRN, in terms of PSNR and SSIM across all datasets. At the $\times2$ scale, our method achieves a PSNR improvement of approximately 0.9 to 1.3 dB over IRN. Importantly, these improvements are not limited to the Y-channel; our method also shows a notable increase in RGB-PSNR by 0.6 to 1.1 dB across the five benchmark datasets. This improvement highlights that our training objective, which is consistent with IRN's and employs similarity loss in the RGB domain, ensures comprehensive enhancement across the full color spectrum, avoiding any bias towards Y-channel optimization. At the $\times4$ scale, although some datasets do not fully surpass the metrics of the original IRN, our method consistently demonstrates superior performance across most datasets. Notably, T-IRN achieves these gains while maintaining a similar parameter count to IRN, highlighting its enhanced efficiency and effectiveness without increasing model complexity.

\paragraph{Ablation Study.}
To assess the effectiveness of T-IRN, we performed an ablation study at the $\times$2 scale, keeping all other parameters constant. (1) We modified IRN to process low-frequency information after converting RGB to YCbCr, without incorporating the T-InvBlock, to compare the V-InvBlock's treatment of luminance and chrominance interactions with our approach. (2) We also tested IRN using an all-zero tensor for high-frequency information to evaluate this strategy's impact. As shown in Table~\ref{tab:ablaction_study_TIRN}, simply converting to YCbCr without the T-InvBlock did not significantly enhance performance, highlighting the V-InvBlock's limitations. In contrast, the full implementation of the T-InvBlock demonstrated clear improvements, confirming its effectiveness. Additionally, the all-zero tensor strategy facilitated better upscaling, affirming the efficacy of both the T-InvBlock and its associated approach in T-IRN.

\paragraph{Qualitative Evaluation.}
\begin{figure*}[t]
    \centering
    \includegraphics[width=\linewidth]{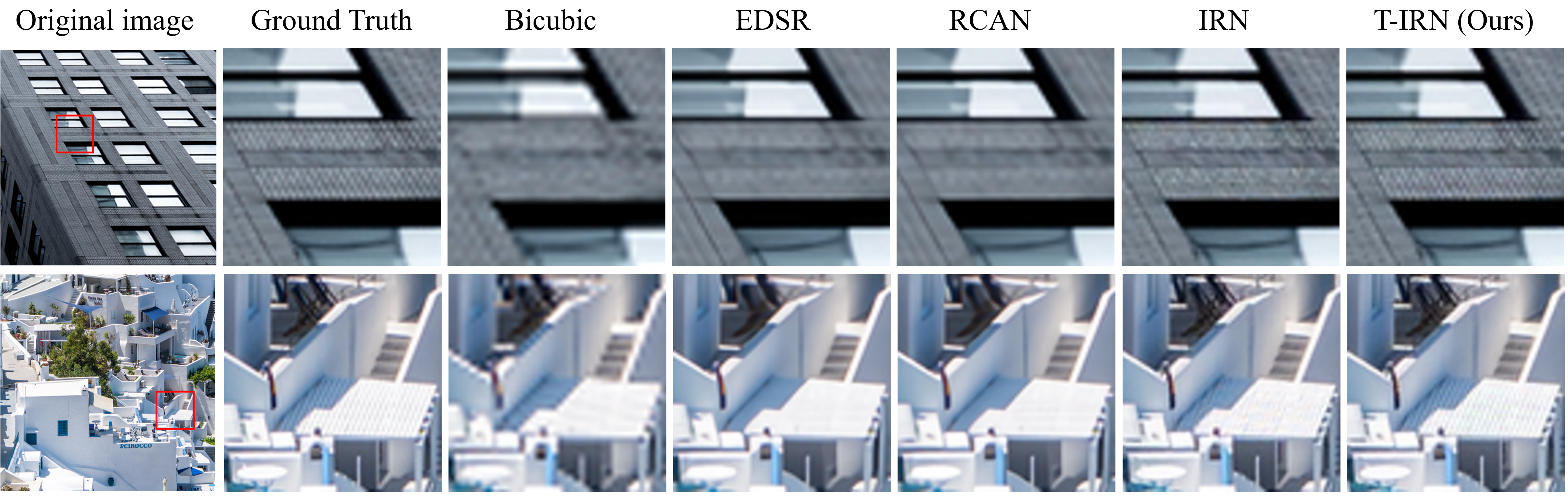}
    \caption{Qualitative Evaluation($\times 2$) for T-IRN. The shown images are 0846 and 0823 from DIV2K.}
    \label{fig:Qualitative_TIRN}
\end{figure*}
\begin{figure*}[t]
    \centering
    \includegraphics[width=\linewidth]{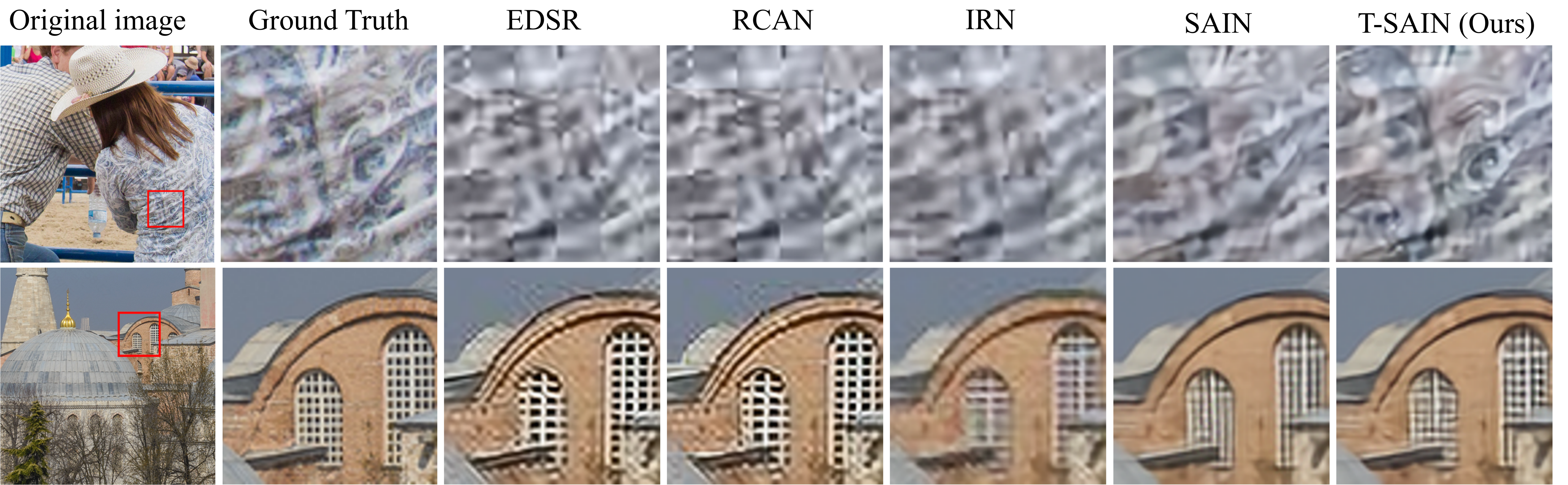}
    \caption{Qualitative Evaluation($\times 2$) for T-SAIN. The shown images are 0804 and 0890 from DIV2K. The first row shows images where the LR input was compressed with JPEG QF=30 before upscaling, while the second row shows images where the LR input was compressed with JPEG QF=90.}
    \label{fig:Qualitative_TSAIN}
\end{figure*}

Figure~\ref{fig:Qualitative_TIRN} shows the visual comparison of HR reconstruction results at $\times 2$ scale between T-IRN and other methods. It is evident that T-IRN consistently recovers more intricate details in the HR images compared to other methods, including IRN.
\subsection{Evaluation for T-SAIN}
\paragraph{Quantitative Evaluation.}
Similarly, we evaluated T-SAIN against SAIN and several other methods previously mentioned, focusing on scenarios where the LR images undergo JPEG lossy compression with QF ranging from 30 to 90. The objective quality of the reconstructed HR images was assessed specifically on the DIV2K dataset.

As shown in Table~\ref{tab:quantitative results T-SAIN}, T-SAIN consistently outperforms SAIN model across various JPEG QF levels on the DIV2K dataset. At the $\times$2 scale, T-SAIN shows PSNR gains of 0.4 to 0.6 dB, along with improved SSIM, across different QF settings. At the $\times$4 scale, it continues to surpass SAIN, with PSNR improvements of 0.2 to 0.5 dB. These results highlight T-SAIN's enhanced robustness and effectiveness in lossy compression scenarios, all while maintaining a similar parameter count to SAIN.

\paragraph{Ablation Study.}
To evaluate T-SAIN's effectiveness, we performed ablation experiments under JPEG compression with QF values of 50, 70, and 80. (1) We applied YCbCr color space conversion to SAIN’s low-frequency branch to assess the model’s handling of luminance and chrominance information. (2) We replaced SAIN's original GMM-based high-frequency modeling with an all-zero tensor to examine the impact of this simplification.

\begin{table}[t]
    \centering
    \small
    \setlength{\tabcolsep}{1mm}
    \begin{tabular}{l|c|c|c|c}
        \hline
        Strategy &Param&QF=50 &QF=70 &QF=80 \\
        \hline
        SAIN &6.49M& 29.05 & 29.83 & 30.13 \\
        \hline
        SAIN (YCbCr) &6.49M& 29.28 & 30.13 & 30.47 \\
        \hline
        SAIN (all-zero) &6.49M& 29.15 & 29.87 & 30.12 \\
        \hline
        T-SAIN&5.45M& \textbf{29.43} & \textbf{30.34} & \textbf{30.69} \\
        \hline
    \end{tabular}
    \label{tab:ablaction_study_T SAIN}
    \caption{Ablation study on T-SAIN: PSNR of different image rescaling strategies for $\times 4$ image reconstruction on DIV2K under QF=50, 70, and 80.}
\end{table}

The results provide key insights. First, converting to YCbCr consistently improved PSNR at all QF levels, probably due to better alignment with JPEG’s YCbCr-based compression, which enhanced the effectiveness of the compression simulator $g$ and improved overall reconstruction quality. Second, replacing GMM modeling with an all-zero tensor improved PSNR at QF=50 and QF=70, with only a slight decline at QF=80. This indicates that concentrating high-frequency information within the LR representation can be more effective for maintaining image quality during reconstruction. Overall, our architectural adjustments, including the tri-branch design, color space conversion, and all-zero mapping strategy, collectively enhanced performance and robustness across compression scenarios.

\paragraph{Qualitative Evaluation.}
Figure~\ref{fig:Qualitative_TSAIN} shows a qualitative comparison of T-SAIN with other methods for $\times 2$ HR reconstruction on JPEG-compressed LR images. T-SAIN clearly recovers more original details compared to other methods, notably SAIN, enhancing compression-awareness and model robustness.

\section{Conclusion}
In this work, we introduced the tri-branch invertible block (T-InvBlock), a novel architecture for enhancing image rescaling. By separating the low-frequency branch into luminance (Y) and chrominance (CbCr) components, T-InvBlock enables more precise processing and improves reconstruction. Our all-zero mapping strategy for high-frequency components further boosts performance. The plug-and-play design of T-InvBlock integrates seamlessly into existing models, as demonstrated in T-IRN and T-SAIN, which outperform traditional methods in both general rescaling and lossy compression scenarios. Evaluations highlight T-InvBlock’s significant advancement in image processing.

\section{Acknowledgement}
This work was supported in part by the National Natural Science Foundation of China under Grant U20A20184, in part by the Natural Science Foundation of Sichuan Province under Grant 2023NSFSC1972.
\bibliography{aaai25}
\end{document}